\documentclass[
 aps,
 prc,
 reprint,
 superscriptaddress,
 nofootinbib,
 longbibliography,
 asmmath,
 amssymb
]{revtex4-2}

\usepackage{amsmath,amssymb,amsfonts,bm}
\usepackage{graphicx}
\usepackage{xcolor}
\usepackage{hyperref}
\usepackage{url}
\usepackage{ragged2e}
\usepackage{array}
\usepackage{booktabs}
\usepackage{multirow}
\usepackage{tikz}
\usepackage{tikz-feynman}
\usepackage{ragged2e}
\usepackage{amsmath}
\usepackage{bm}
\usepackage{array}
\usepackage{caption}
\usepackage{slashed}

\makeatletter
\renewcommand{\fnum@figure}{Figure~\thefigure}
\makeatother

\begin{document}

\title{The leading nuclear-structure electrostatic correction in arbitrary $\beta$ decays}

\author{Daniel Benatar}
\email{daniel.benatar@mail.huji.ac.il}
\affiliation{Racah Institute of Physics, The Hebrew University, The Edmond J. Safra Campus Givat Ram, Jerusalem 9190401, Israel}

\author{Ayala Glick-Magid}
\email{ayalaglick@tauex.tau.ac.il}
\affiliation{School of Physics and Astronomy, Tel-Aviv University, Tel-Aviv 69978, Israel}

\author{Doron Gazit}
\email{doron.gazitr@mail.huji.ac.il}
\affiliation{Racah Institute of Physics, The Hebrew University, The Edmond J. Safra Campus Givat Ram, Jerusalem 9190401, Israel}

\date{June 6, 2026}

\begin{abstract}
We develop a systematic theoretical framework to improve theoretical predictions for nuclear $\beta$ decays of arbitrary angular momentum $J$, leading to a model-independent nuclear-structure electrostatic correction to the Coulomb interaction between the emitted lepton and the nuclear charge distribution, useful for ongoing and future precision searches for physics beyond the Standard Model.  
The formalism is based on nuclear matrix elements expanded in multipole operators, as commonly used in \emph{ab initio} calculations. 
First-order Coulomb corrections are derived from one-photon exchange 
preserving the full multipole and angular structure of the decay rate, and are subsequently expanded in the relevant small parameters of the nuclear problem, suppressing the leading nuclear structure correction to a few per-mills for medium mass nuclei with natural beta decay properties.
We show that within this formalism, the leading Coulomb correction originates from three modifications of the original weak-only interaction: a modification of the nuclear charge form factor, which yields a correction similar to the known Fermi function, a shift of the momentum transfer within the lepton traces, and the same shift but inside the nuclear multipole operators.
We additionally provide explicit results for allowed Gamow--Teller and unique first-forbidden transitions.

\end{abstract}


\maketitle

\section{\textbf{Introduction}}
In the Standard Model (SM) \cite{Langacker2010, GrossmanNirSM2022}, $\beta$ decay is a fundamental weak-interaction process in which a quark undergoes a flavor-changing transition mediated by $\mathcal{W}$-boson exchange, accompanied by the emission of a charged lepton and a (anti)neutrino \cite{BehrensBuehring1982, BehrensBogdan1970, BuehringScholz1965, BarrNPP2016}. These processes convert a neutron into a proton, or vice versa \cite{Herczeg2001}. At the nucleon level, they are described as
\[
\begin{aligned}
\text{$\beta^-$ decay} &: \quad n \rightarrow p + e^- + \bar{\nu}_e \;, \\
\text{$\beta^+$ decay} &: \quad p \rightarrow n + e^+ + \nu_e \;.
\end{aligned}
\]

Apart from the weak interaction, the emitted charged lepton also undergoes a Coulomb interaction with the nuclear charge distribution. This interaction induces corrections that scale with $\alpha Z$, where $\alpha \simeq 1/137$ and $Z$ is the nuclear charge. While typically small, these effects become increasingly important in high-precision $\beta$-decay studies, particularly in searches for physics beyond the SM (BSM), where sub-percent-level accuracy is required. A systematic treatment is therefore essential for meaningful comparison with experimental data \cite{Kleesiek2019, Brodeur2023NuclearBetaDecayBSM, Falkowski_2021, GlickMagid2022}.

The theoretical description is based on the V--A structure of the charged weak interaction and the semileptonic current--current formalism \cite{BehrensBuehring1982, Walecka2004}.
In the standard treatment, Coulomb effects are incorporated through the Fermi function \cite{Morita1963, Bottino1973, HillPlestid2024Field}, which describes the propagation of the charged lepton in the nuclear Coulomb field. This function is obtained by solving the Dirac equation for a point-like charge. Extensions to realistic finite nuclear charge distributions have been extensively studied, with particular emphasis on allowed channels (see, e.g., Ref.~\cite{Hayen2018}). Extensions to forbidden transitions exist, in particular in the seminal work by Behrens and B\"uhring \cite{BehrensBuehring1982}. However, these depend on model choice for the charge density of the nucleus, usually a uniform charge density, limiting controlled estimates of theoretical uncertainties, and are formulated in ways that are challenging to incorporate in modern calculations.

In this work, we develop a systematic and model-independent theoretical framework for incorporating the leading nuclear-structure Coulomb (electrostatic) correction to arbitrary $\beta$ transitions. The dimensionless parameter that governs the correction scales as
\begin{equation}
\label{eq:psilon_EC}
\varepsilon_{EC}=\alpha Z\,E_e R   \approx 4\cdot 10^{-5} \left(\frac{E_e}{1\text{\,MeV}}\right)ZA^{1/3},
\end{equation}
where $E_e$ is the charged-lepton energy and $ R$ characterizes the nuclear radius. The numerical estimate uses a typical $\beta$-decay lepton energy $E_e \sim 1\,\mathrm{MeV}$ and the phenomenological relation $R\approx 1.2 A^{1/3}\,\mathrm{fm}$\footnote{Natural units $\hbar=c=1$ are assumed throughout the paper}, which places the correction within the sensitivity of current precision experiments.

This formulation enables a separation between the point-nucleus contribution, encoded in the Fermi function, and the finite-size nuclear correction. Higher-order contributions are expected at orders $(\alpha Z)\cdot \varepsilon_{EC}$ and $(E_e R)\cdot \varepsilon_{EC} $. In addition, the formulation is readily compatible with standard \emph{ab initio} calculations. It applies to $\beta$ decays of arbitrary angular momentum $J$, including both allowed and forbidden transitions. Our work complements previous studies in which electrostatic corrections were not included \cite{GlickMagid2022, GlickMagid2023}.

The paper is structured as follows. In Section~\ref{sec_2}, following Holstein \cite{Holstein1974}, the free lepton wave function $\psi_e(\mathbf{r})$ is replaced with the Coulomb-distorted wave function $\phi_e(\mathbf{r})$, obtained from the Dirac equation in the electrostatic potential $V(\mathbf{r})$ of the nucleus\cite{Holstein1974, Bottino1972, PeskinSchroeder1995, Holstein1983}. This modifies the weak Hamiltonian $\hat{\mathcal H}_w$ \cite{BehrensBuehring1982, Bambynek1977} through a Coulomb-corrected leptonic current $L_\mu^{c}$ \cite{BehrensBuehring1982, Walecka2004, Hayen2018, Holstein1974}, which yields the modified matrix element
\begin{equation}
\langle f | \hat{\mathcal H}_w^{c} (\boldsymbol{r}) | i \rangle
=
\langle f | \hat{\mathcal H}_w (\boldsymbol{r})| i \rangle
+
\langle f | \hat{\mathcal H}^{c} (\boldsymbol{r}) | i \rangle \;,
\label{eq:Hw_total}
\end{equation}
where $\hat{\mathcal H}^{c}$ arises from one-photon exchange between the emitted lepton and the nuclear charge distribution. The decay rate $dw_{\beta^\mp}^{c}$ is obtained from
$\bigl|\langle f | \hat{\mathcal H}_w^{c}(\boldsymbol r) | i \rangle\bigr|^{2}$.

The Coulomb interaction modifies the leptonic wave function and induces a shift in the momentum transfer, whose first-order expansion is discussed in Section~\ref{sec_3}.
 Integration over the exchanged-photon momentum, together with the nuclear charge form factor, contains a point-nucleus contribution that recovers the Fermi function $F(Z,E_e)$ \cite{Walecka2004, GlickMagid2017}. In addition, integration over the exchanged-photon momentum, combined with the traces of the Coulomb-corrected leptonic current and the Coulomb-induced shift of the multipole operators within the multipole expansion
\cite{Walecka2004,GlickMagid2022,DonnellyHaxton1979,DonnellyWalecka1976,Ryder1996,Walecka2010Advanced,Walecka2001},
generates an angular-dependent finite-size nuclear-structure correction of order $\varepsilon_{EC}$ beyond the point-charge approximation.

We conclude by deriving a general expression for the $\beta$-decay rate for arbitrary nuclear transitions, including both the point-nucleus contribution and the finite-size correction of order $\varepsilon_{EC}$, and by presenting representative applications to Gamow--Teller and unique first-forbidden decays, of relevance for precision tests of the Standard Model and searches for BSM physics.

\section{\textbf{Theoretical description of electrostatically corrected $\beta$-decay in Nuclei}}
\label{sec_2}
In this section, we incorporate the Coulomb-interaction corrections into the $\beta$-decay rate by employing Coulomb-distorted lepton wave functions and constructing the corresponding matrix elements.

\subsection{Electrostatically distorted lepton wave function}
We begin by introducing the distorted lepton wave functions, $\phi_{e^{\mp}}(\mathbf{r})$, defined as the solutions of the Dirac equation in the presence of the nuclear Coulomb potential $V_{\beta^{\mp}}(r)$, incorporating the lowest-order Coulomb correction, which corresponds to one-photon exchange, as shown in Figure~\ref{fig:beta_decay} \cite{Hayen2018, Bottino1973, Holstein1974, Bottino1974, Beg1972, HillPlestid2024all},
\begin{align}
\overline{\phi}_{e^-}(r)
&=
\overline{u}_e(k)e^{-i\boldsymbol{k}\cdot\boldsymbol{r}}
\nonumber\\
&\quad
-i\int d^4z\,
\overline{u}_e(k)e^{ik\cdot z}
\gamma_0 V_{\beta^-}(r)\mathcal{S}_F(z-r) ,
\label{eq:elec}
\\
\overline{\phi}_{e^+}(r)
&=
\overline{v}_e(k)e^{-i\boldsymbol{k}\cdot\boldsymbol{r}}
\nonumber\\
&\quad
-i\int d^4z\,
\overline{v}_e(k)e^{ik\cdot z}
\gamma_0 V_{\beta^+}(r) S_F(z-r) .
\label{eq:pos}
\end{align}
Here, $u_e(k)$ ($v_e(k)$) denotes the outgoing electron (positron) spinor for $\beta^-$ ($\beta^+$) decay, $k$ is the corresponding charged-lepton four-momentum, and the lepton Feynman propagator is~\cite{PeskinSchroeder1995}:
\begin{equation}
        \mathcal{S}_F (z-r) = \int \dfrac{d^4k}{(2\pi)^4} \dfrac{i (\slashed{k} + m_e)}{k^2-m_e^2+i\epsilon} e^{-ik \cdot(z-r)},
\end{equation}
with $m_e$ the mass of the electron.

We follow the approach of Holstein and express the electrostatic potential for $\beta^{\mp}$ decay in terms of the nuclear charge form factor $\mathcal{F}(k_\gamma^2)$~\cite{Bottino1973,Holstein1974,ArmstrongKim1972,Bottino1974}:
\begin{equation}
V_{\beta^\mp}(r)
=
\mp\,8\pi Z\alpha
\int \frac{d^4 k_\gamma}{(2\pi)^4}
\frac{\delta(k_{\gamma 0})}
{k_\gamma^2+i\epsilon}
e^{ik_\gamma \cdot r}
\mathcal{F}(k_\gamma^2) ,
\end{equation}
where $k_{\gamma}$ denotes the four-momentum carried by the exchanged virtual photon, and $k_{\gamma 0}$ is its energy.
\begin{figure}[t]
\centering
\begin{tikzpicture}

  \begin{scope}[shift={(-4,0)}]

    \draw[fill=gray!20] (0,0) ellipse (0.5 and 0.3);

    \draw[double, very thick] (-2,-1) -- (0,0);
    \node at (-2.3,-1.3)
    {$\boldsymbol{^{A}_{Z}X}$};

    \draw[->, double, very thick] (0,0) -- (2,-1);
    \node at (2.7,-1.3)
    {$\boldsymbol{^{A}_{Z+1}X'}$};

    \draw[
      thick,
      postaction={
        decorate,
        decoration={
          markings,
          mark=at position 0.5 with {\arrow{>}}
        }
      }
    ]
    (0,0) -- (2,0.5);

    \node at (1.2,0.62)
    {$\boldsymbol{e^-}$};

    \draw[
      decorate,
      decoration={
        snake,
        amplitude=2pt,
        segment length=6pt
      },
      thick
    ]
    ($(0,0)!0.40!(3.5,0.85)$)
    --
    ($(0,0)!0.40!(3.5,-1.9)$)
    node[midway,right]
    {$\gamma$};

    \draw[
      thick,
      postaction={
        decorate,
        decoration={
          markings,
          mark=at position 0.5 with {\arrow{<}}
        }
      }
    ]
    (0,0) -- (1.5,1.5);

    \node at (0.45,1.0)
    {$\boldsymbol{\overline{\nu}_e}$};

    \node at (1.95,1.7)
    {$\boldsymbol{\nu}$};

    \node at (2.45,0.62)
    {$\boldsymbol{k}$};

    \node at (0,-2.1)
    {\small (a) $\beta^-$ decay};

  \end{scope}

  \begin{scope}[shift={(-4,-4.5)}]

    \draw[fill=gray!20] (0,0) ellipse (0.5 and 0.3);

    \draw[double, very thick] (-2,-1) -- (0,0);
    \node at (-2.3,-1.3)
    {$\boldsymbol{^{A}_{Z}X}$};

    \draw[->, double, very thick] (0,0) -- (2,-1);
    \node at (2.7,-1.3)
    {$\boldsymbol{^{A}_{Z-1}X'}$};

    \draw[
      thick,
      postaction={
        decorate,
        decoration={
          markings,
          mark=at position 0.5 with {\arrow{<}}
        }
      }
    ]
    (0,0) -- (2,0.4);

    \node at (1.2,0.58)
    {$\boldsymbol{e^+}$};

    \draw[
      decorate,
      decoration={
        snake,
        amplitude=2pt,
        segment length=6pt
      },
      thick
    ]
    ($(0,0)!0.40!(3.6,0.8)$)
    --
    ($(0,0)!0.40!(3.55,-1.9)$)
    node[midway,right]
    {$\gamma$};

    \draw[
      thick,
      postaction={
        decorate,
        decoration={
          markings,
          mark=at position 0.5 with {\arrow{>}}
        }
      }
    ]
    (0,0) -- (1.5,1.5);

    \node at (0.45,1.0)
    {$\boldsymbol{\nu_e}$};

    \node at (1.95,1.7)
    {$\boldsymbol{\nu}$};

    \node at (2.45,0.58)
    {$\boldsymbol{k}$};

    \node at (0,-2.1)
    {\small (b) $\beta^+$ decay};

  \end{scope}

\end{tikzpicture}

\caption{
Nuclear $\beta^\mp$ decay with the Coulomb interaction represented by one-photon exchange between the emitted charged lepton and the nuclear charge distribution. 
$X$ and $X'$ denote the initial and final nuclear states, with mass number $A$ and charge $Z$.
}
\label{fig:beta_decay}

\end{figure}
Revisiting Eqs.~\eqref{eq:elec} and~\eqref{eq:pos}, we substitute these expressions into the distorted wave functions, obtaining
\begin{align}
    \overline{\phi}_{e^-}(r)
    &=
    \overline{u}_e(k)e^{-i\boldsymbol{k}\cdot\boldsymbol{r}}
    +i\int d^4z\,
    \overline{u}_e(k)e^{ik\cdot z}\gamma_0\,8\pi Z\alpha
    \nonumber\\
    &\quad\times
    \int \frac{d^4k_\gamma}{(2\pi)^4}
    \frac{\delta(k_{\gamma 0})}{k_\gamma^2+i\epsilon}
    e^{ik_\gamma\cdot r}\mathcal{F}(k_\gamma^2)
    \nonumber\\
    &\quad\times
    \int \frac{d^4k'}{(2\pi)^4}
    \frac{i(\slashed{k'}+m_e)}{k'^2-m_e^2+i\epsilon}
    e^{-ik'\cdot(z-r)} ,\\
    \overline{\phi}_{e^+}(r)
    &=
    \overline{v}_e(k)e^{-i\boldsymbol{k}\cdot\boldsymbol{r}}
    -i\int d^4z\,
    \overline{v}_e(k)e^{ik\cdot z}\gamma_0\,8\pi Z\alpha
    \nonumber\\
    &\quad\times
    \int \frac{d^4k_\gamma}{(2\pi)^4}
    \frac{\delta(k_{\gamma 0})}{k_\gamma^2+i\epsilon}
    e^{ik_\gamma \cdot r}\mathcal{F}(k_\gamma^2)
    \nonumber\\
    &\quad\times
    \int \frac{d^4k'}{(2\pi)^4}
    \frac{i(\slashed{k'}+m_e)}{k'^2-m_e^2+i\epsilon}
    e^{-ik'\cdot(z-r)} ,
\end{align}
where $\nu$ is the four-momentum of the (anti)neutrino.

The nuclear charge form factor $\mathcal{F}(k_\gamma^2)$ is defined as the Fourier transform of the charge distribution \cite{Bottino1973}. Motivated by the application of this framework to future precision $\beta$-decay experiments, where uncertainty estimates are essential, and noting that the momentum transfer in $\beta$ decay is small compared to the wave number of the nucleus, i.e., to $1/R$, we express the form factor in a convenient form that is exact to $\mathcal{O}(k_\gamma^2)$~\cite{Bottino1973}:
\begin{equation} \label{eq:FF}
    \mathcal{F}(k_\gamma^2) = \dfrac{1}{1+a^2k_\gamma^2},
\end{equation}
with $a=\dfrac{1}{\sqrt{6}} \langle r^2 \rangle^{1/2}$, where $\langle r^{2} \rangle$ is the nuclear mean-square charge radius of the final nucleus, which can be taken from experiment or calculated using \emph{ab initio} or other nuclear models. One of the advantages of the present work is that the dependence on nuclear structure enters only through the nuclear charge radius, making the correction largely model independent.

An essential observation, made by Holstein~\cite{Holstein1974}, is that the exchanged Coulomb photon momentum is soft compared to the typical momentum transfer in the decay, stemming from the small recoil of the heavy nucleus. Additionally, the form factor induces finite-size suppression, resulting in $k_\gamma/k \sim 1/(MR) \ll 1$, where $M$ denotes the nuclear mass, thus expressing the hierarchy between nuclear and leptonic momentum scales. Using $M \simeq A m_N$ with the previously mentioned phenomenological relation for $R$, one finds
\begin{equation}
MR \simeq (m_N r_0)\,A^{4/3} \approx 5.7\,A^{4/3} \gg 1 \, .
\end{equation}
This directly implies
$k_\gamma \ll k$ for all nuclei \cite{Holstein1974}, yielding the final expression for the distorted lepton wave function:
\begin{multline}
       \overline{\phi}_{e^-}(\boldsymbol{r}) = \overline{u}(k) e^{-i\boldsymbol{k} \cdot\boldsymbol{r}} - \gamma_0 (\slashed{k} + m_e) \overline{u}(k)\\
       \times
       \int_{0}^{\infty} dk_\gamma\,k_\gamma^{2}\ e^{-i(\boldsymbol{k} - \boldsymbol{k}_\gamma)\cdot \boldsymbol{r}} \dfrac{1}{(2\pi)^3} \int d\Omega_{\gamma} \Theta(k,k_\gamma),
\end{multline}
\begin{multline}
    \overline{\phi}_{e^+}(\boldsymbol r) =
\overline{v}(k)e^{-i\boldsymbol{k}\cdot\boldsymbol r}
+
\overline{v}(k)\gamma_0(\slashed{k}+m_e)\\
\times
\int_0^\infty dk_\gamma\,k_\gamma^2\,
e^{-i(\boldsymbol{k}-\boldsymbol{k}_\gamma)\cdot\boldsymbol r} \dfrac{1}{(2\pi)^3} \int d\Omega_{\gamma} 
\Theta(k,k_\gamma),
\end{multline}
where we introduce the Coulomb kernel,
\begin{equation}
\Theta(k,k_\gamma)
\equiv 8\pi\alpha Z
\frac{\mathcal{F}(k_\gamma^{2})}
{\big(k_\gamma^{2}+i\epsilon\big)\,
 \big(k_\gamma^{2}-2 k k_\gamma\cos\theta + i\epsilon\big)},
 \label{eq:Theta}
\end{equation}
and
$d\Omega_{\gamma}
\equiv
\sin\theta\, d\theta\, d\phi$
denotes the solid-angle element associated with the exchanged-photon momentum $\boldsymbol{k}_{\gamma}$, with
$0\le\theta\le\pi$
and
$0\le\phi\le2\pi.$
\subsection{Transition matrix element with distorted lepton wave functions}
We next use the distorted lepton wave function to calculate the
transition matrix element. The effective weak Hamiltonian governing
low-energy processes in the SM, including the Coulomb
interaction, is given by Refs.~\cite{Walecka2004, ArmstrongKim1972, Bambynek1977}:
\begin{equation}
   \mathcal{H}^c_w (\boldsymbol{r})= -\dfrac{G_F V_{ud}}{\sqrt{2}} \Big[ \mathcal{J}^{\mu}(\boldsymbol{r}) j^c_\mu(\boldsymbol{r})  + h.c. \Big],
\end{equation}
where the Fermi coupling constant is $G_F = 1.1663787\times10^{-5} \mathrm{GeV}^{-2}$, and $V_{ud} = 0.97370$ is the up--down element of the Cabibbo--Kobayashi--Maskawa (CKM) quark-mixing matrix~\cite{huertas2003effectivefieldtheoryapproach}.

We employ the distorted lepton wave function
$\overline{\phi}_e(\boldsymbol{r})$ in the leptonic current, thereby obtaining the Coulomb-corrected leptonic current:
\begin{equation}
    j^c_\mu(\boldsymbol r)
    =\overline{\phi}_e(\boldsymbol r)\,
    \gamma_\mu(1+\gamma_5)\,
    \psi_{\nu_e}(\boldsymbol r)\, .
\end{equation}
Here $\mathcal{J}^\mu(\boldsymbol{r})$ and $j^c_\mu(\boldsymbol{r})$ denote the hadronic and leptonic currents, respectively, and
$\psi_{\nu_e}(\boldsymbol{r}) = v_\nu e^{-i\boldsymbol{\nu}\cdot \boldsymbol{r}}$ is the (anti)neutrino spinor wave function. 

The corresponding matrix element, including the Coulomb interaction between the initial $(i)$ and final $(f)$ nuclear states, is given by:
\begin{equation}
    \langle f \vert \mathcal{\hat{H}}^c_{w} (\boldsymbol{r}) \vert i \rangle =
    -\frac{G_F V_{ud}}{\sqrt{2}}
    \int d^3r\,
    \langle f \vert
    \hat{\mathcal{J}}^{\mu}(\boldsymbol{r})
    j^c_{\mu}(\boldsymbol{r})
    \vert i \rangle ,
\end{equation}
so the generalized matrix element incorporating the Coulomb-interaction correction, with $\boldsymbol{q}=\boldsymbol{k}+\boldsymbol{\nu}$ the original momentum transfer,
and
$\boldsymbol{q}'= \boldsymbol{k} + \boldsymbol{\nu} - \boldsymbol{k}_{\gamma}$ the momentum transfer modified by the exchanged-photon momentum,
is given by
\begin{multline}
    \langle f \vert \mathcal{\hat{H}}^c_{w} (\boldsymbol{r}) \vert i \rangle =
-\frac{G_F V_{ud}}{\sqrt{2}}
\int d^3r\,
\left\langle f \left|
\hat{\mathcal J}_{\mu}(\boldsymbol r)\,
e^{-i\boldsymbol q\cdot \boldsymbol r}\,
l_\mu^{\mp}
\right| i \right\rangle
\\
\quad
\pm
\frac{G_F V_{ud}}{\sqrt{2}}
\int_0^\infty dk_\gamma\, k_\gamma^2 \dfrac{1}{(2\pi)^3} \int d\Omega_{\gamma} 
\int d^3r\ \\
\quad\times
\left\langle f \left|
\hat{\mathcal J}_{\mu}(\boldsymbol r)\,
e^{-i\boldsymbol q'\cdot \boldsymbol r}\,
\Theta(k,k_\gamma)\,
L_\mu^{c^{\mp}}
\right| i \right\rangle ,
\end{multline}
where the free leptonic currents for $\beta^{\mp}$ decay are
\begin{align}
l_\mu^{-} &= \overline{u}_e(k)\, \gamma_\mu (1 + \gamma_5)\, v_{\nu_e}(\nu),\\
l_\mu ^{+}
&=
\overline{u}_{\nu_e}(\nu)\,
\gamma_\mu(1+\gamma_5)\,
v_e(k),
\end{align}
and the Coulomb-corrected leptonic currents are
\begin{align}
        L_\mu^{c-} &= \overline{u}_{e}(k)\gamma_0 (\slashed{k} + m_e) \gamma_{\mu}(1+\gamma_5) v_{\nu_e}(\nu),\\
L_\mu^{c+} &=
\overline{u}_{\nu_e}(\nu)\,
\gamma_\mu(1+\gamma_5)\,
(\slashed{k}+m_e)\gamma_0\,
v_e(k).
\end{align}

We employ a multipole expansion of both $e^{-i\boldsymbol{q}\cdot\boldsymbol{r}}$ and $e^{-i\boldsymbol{q}'\cdot\boldsymbol{r}}$, along with a corresponding vector multipole expansion, decomposing every three-vector as
  $\boldsymbol{L}=\sum_{\lambda=0,\pm1}L_{\lambda} \boldsymbol{ \mathrm{e}}_{\lambda}^{\dagger}$, following the formalism in~\cite{Walecka2004}.
Subsequently, the Coulomb-corrected matrix element takes the form:
\begin{multline}
    \langle f \vert \mathcal{\hat{H}}^c_{w} (\boldsymbol{r}) \vert i \rangle\\
     = -\frac{G_F V_{ud}}{\sqrt{2}}
    \left\langle f \left|
    \Bigg\{
    -\sum_{J \geq 1} \sqrt{2\pi(2J+1)}(-i)^J
    \nonumber\right.\right.\\
    \times
    \sum_{\lambda=\pm1} l_\lambda^{\mp}
    \Big[\lambda \hat{\mathcal{T}}^{mag}_{J-\lambda}(q)
    +\hat{\mathcal{T}}^{el}_{J-\lambda}(q)\Big]
    \\
    +
    \sum_{J \geq 0}\sqrt{4\pi(2J+1)}(-i)^J
    \Big[l_3^{\mp} \hat{\mathcal{L}}_{J0}(q)
    - l_0^{\mp} \hat{\mathcal{M}}_{J0}(q)\Big]
    \\
    \mp
    \int_0^\infty dk_\gamma\,k_\gamma^2
    \Bigg\{
    -\sum_{J \geq 1}\sqrt{2\pi(2J+1)}(-i)^J
    \\
    \times \sum_{\lambda=\pm1}\Theta(k,k_\gamma)L^{c^{\mp}}_\lambda
    \Big[\lambda \hat{\mathcal{T}}^{mag}_{J-\lambda}(q')
    +\hat{\mathcal{T}}^{el}_{J-\lambda}(q')\Big]
    \\
    +
    \sum_{J \geq 0}\sqrt{4\pi(2J+1)}(-i)^J
    \Big[\Theta(k,k_\gamma)L^{c^{\mp}}_3 \hat{\mathcal{L}}_{J0}(q')
    \\
    -\Theta(k,k_\gamma)L^{c^{\mp}}_0 
    \hat{\mathcal{M}}_{J0}(q')\Big]
    \Bigg\}
    \Bigg\} \Bigg| i \Bigg\rangle ,
\end{multline}
using the standard definition of the Coulomb kernel,
($\hat{\mathcal{M}}_{JM}$), longitudinal 
($\hat{\mathcal{L}}_{JM}$), electric 
($\hat{\mathcal{T}}_{JM}^{\mathrm{el}}$), and magnetic 
($\hat{\mathcal{T}}_{JM}^{\mathrm{mag}}$) multipole operators~\cite{Walecka2004,GlickMagid2023,huertas2003effectivefieldtheoryapproach}:
\begin{align}
    \mathcal{\hat{M}}_{JM}(q) &= \int d^3r \big[j_{J}(q r) \mathcal{Y}_{JM}(\Omega_{r})\big] \mathcal{\hat{J}}_{0}\boldsymbol{(r)},\\
    \mathcal{\hat{L}}_{JM}(q) &= \dfrac{i}{q}\int d^3r \Big\lbrace\nabla \big[j_{J}(q r) \mathcal{Y}_{JM}(\Omega_{r})\big]\Big\rbrace \cdot \boldsymbol{\mathcal{\hat{J}}}(\boldsymbol{r}),\\
    \mathcal{\hat{T}}^{el}_{JM}(q) &= \dfrac{1}{q}\int d^3r \big[\nabla \times j_{J}(q r) \boldsymbol{\mathcal{Y}}^{M}_{JJ1} (\Omega_{r})\big] \cdot \boldsymbol{\mathcal{\hat{J}}}(\boldsymbol{r}),\\
    \mathcal{\hat{T}}^{mag}_{JM}(q) &= \int d^3r \big[ j_{J}(q r) \boldsymbol{\mathcal{Y}}^{M}_{JJ1}(\Omega_{r})\big] \cdot \boldsymbol{\mathcal{\hat{J}}}(\boldsymbol{r}),
\end{align}
evaluated both in momentum transfer $q\equiv |\boldsymbol{q}|$ and in Coulomb-shifted momentum transfer $q'\equiv |\boldsymbol{q}'|$, which incorporates the exchanged-photon momentum.
Here, $J$ denotes the multipole rank (total angular momentum) and $M$ is its magnetic projection. $j_J(x)$ is the spherical Bessel function of order $J$, while $\mathcal{Y}_{JM}$ and $\boldsymbol{\mathcal{Y}}_{JL1}^{M}$ denote the spherical and vector spherical harmonics, respectively.

We compute the $\beta$-decay matrix element amplitude as a preliminary step toward the derivation of the full decay rate. To first order in the Coulomb interaction, the $\mathcal{O}(\alpha Z)$ contribution arises exclusively from the interference term proportional to $\Theta(k,k_\gamma)$:
\begin{multline} \label{eq:H^C}
    \Big|\langle f \vert \mathcal{\hat{H}}^c_{w} (\boldsymbol{r}) \vert i \rangle \Big|^{2} =
    \frac{G_{F}^{2}\,|V_{ud}|^{2}}{2}
    \Bigg[
    \Big|\langle f \vert \mathcal{\hat{H}}_{w} (\boldsymbol{r}) \vert i \rangle \Big|^{2} \\
     \pm
    2 \mathcal{R}e \langle f \vert \mathcal{\hat{H}}_{w} (\boldsymbol{r}) \vert i \rangle\ \langle f \vert \hat{\mathcal{H}}^c (\boldsymbol{r}) \vert i \rangle^{*}
    \Bigg]  +\mathcal{O}\big((\alpha Z)^2\big),
\end{multline}
where
\begin{multline}
 \langle f \vert \mathcal{\hat{H}}_{w} (\boldsymbol{r}) \vert i \rangle 
 = -\frac{G_F V_{ud}}{\sqrt{2}} \bigg[ - \sum_{J \ge 1} \sqrt{2\pi(2J+1)}(-i)^{J} \\
\times
        \sum_{\lambda = \pm 1} l^{\mp}_{\lambda} 
        \Big\langle f\Big|\lambda \,\hat{\mathcal{T}}^{\mathrm{mag}}_{J-\lambda}(q) +   \hat{\mathcal{T}}^{\mathrm{el}}_{J-\lambda}(q) \Big| i \Big\rangle \\
 + \sum_{J \ge 0}\sqrt{4\pi(2J+1)} (-i)^{J} \\
\times
        \Big[
        l_{3}^{\mp}\,\big\langle f\big|\hat{\mathcal{L}}_{J0}(q)\big|i\big\rangle 
       -l_{0}^{\mp}\,\big\langle f\big|\hat{\mathcal{M}}_{J0}(q)\Big|i\Big\rangle
       \Big] \bigg]
\end{multline}
is the standard matrix element, and
\begin{multline}
    \langle f \vert \hat{\mathcal{H}}^c (\boldsymbol{r})\vert i \rangle =
    \mp \frac{G_F V_{ud}}{\sqrt{2}} \Bigg\{ \int_{0}^{\infty} dk_\gamma\,k_\gamma^{2} \dfrac{1}{(2\pi)^3} \int d\Omega_{\gamma} \\
    \quad\times
     \bigg[
    - \sum_{J \ge 1} \sqrt{2\pi(2J+1)}(-i)^{J}
    \\
    \quad\times
    \sum_{\lambda = \pm 1}
    \Theta(k,k_\gamma)\,L^{c^{\mp}}_{\lambda}
    \Big\langle f\Big|
    \lambda \hat{\mathcal{T}}^{\mathrm{mag}}_{J-\lambda}(q')
    + \hat{\mathcal{T}}^{\mathrm{el}}_{J-\lambda}(q')
    \Big| i \Big\rangle
    \\
    \quad+
    \sum_{J \ge 0}\sqrt{4\pi(2J+1)}(-i)^{J}
    \Big[
    \Theta(k,k_\gamma)\,L^{c^{\mp}}_{3}
    \big\langle f\big|\hat{\mathcal{L}}_{J0}(q')\big|i\big\rangle
    \\
    \qquad
    -\Theta(k,k_\gamma)\,L^{c^{\mp}}_{0}
    \Big\langle f\Big|\hat{\mathcal{M}}_{J0}(q')\big|i\big\rangle
    \Big] \bigg]
    \Bigg\}
\end{multline}
is the Coulomb-induced matrix element.  For convenience, we will suppress the $\mp$ labels on the leptonic currents in the following.

Turning to the standard expression for the $\beta$-decay rate~\cite{Walecka2004},
\begin{multline}
            dw^c_{\beta^{\mp}} = \dfrac{1}{(2\pi)^5}(E_0 -E_e) k E_e dE_e d\Omega_k d\Omega_\nu \\
        \times \sum_{\text{lepton spins}} \dfrac{1}{(2J_i+1)}\sum_{M_i} \sum_{M_f} \Big|\Big\langle f \vert \mathcal{\hat{H}}^c_{w} (\boldsymbol{r}) \vert i \Big\rangle \Big|^{2},
\end{multline}
we substitute Eq.~\eqref{eq:H^C}, applying the Wigner--Eckart theorem and introducing reduced matrix elements.
The standard leptonic traces associated with the regular weak interaction, $1/2 \sum_{\text{lepton spins}} l_{\mu} l_{\mu'}$, are well known~\cite{Walecka2004}.
Interestingly, the traces associated with the Coulomb-induced leptonic structures in the $\beta$-decay matrix element,
$\frac{1}{2}\sum_{\text{lepton spins}}
l_\mu L_{\mu'}^{c*}$,
are found to retain the same structure as the corresponding traces in the absence of the Coulomb interaction, differing only by an overall factor of $2E_e$ and by momentum replacement
$q\rightarrow q'$.

Finally, employing the Coulomb-corrected leptonic current and inserting the corresponding leptonic traces together with the explicit multipole operators, we obtain the Coulomb-corrected $\beta$-decay rate:
\begin{align}
dw^c_{\beta^{\mp}}
&=
\dfrac{G_F^2 |V_{ud}|^2}{2\pi^3}
(E_0-E_e)^2 k E_e\, dE_e\,
\dfrac{d\Omega_k}{4\pi}
\dfrac{d\Omega_{\nu}}{4\pi}
\dfrac{4\pi}{2J_i+1}\nonumber\\
&\quad\times
\bigg\{
\mathcal{M}^{0}
+
\mathcal{M}^{c}
\bigg\},
\label{eq:beta}
\end{align}
where the first term corresponds to the standard $\beta$-decay multipole expansion~\cite{Walecka2004}:
\begin{align}
\mathcal{M}^{0}
&=
\sum_{J \geq 1} \bigg[
\Big(1-( \boldsymbol{\hat{\nu}} \cdot \boldsymbol{\hat{q}} )
( \boldsymbol{\beta} \cdot \boldsymbol{\hat{q}} )\Big)
\Big(
\vert \langle J_f \Vert \mathcal{\hat{T}}_J^{mag}(q)\Vert J_i\rangle \vert^2
\nonumber\\
&+
\vert \langle J_f \Vert \mathcal{\hat{T}}_J^{el}(q)\Vert J_i\rangle \vert^2
\Big)
\pm
\boldsymbol{\hat{q}}\cdot ( \boldsymbol{\hat{\nu}}-\boldsymbol{\beta})
\nonumber\\
&\times
2 \mathcal{R}e\,\Big(
\langle J_f \Vert \mathcal{\hat{T}}_J^{mag}(q)\Vert J_i\rangle
\langle J_f \Vert \mathcal{\hat{T}}_J^{el}(q)\Vert J_i\rangle^*
\Big)
\bigg]
\nonumber\\
&+
\sum_{J \geq 0} \bigg[
\Big(
1-\boldsymbol{\hat{\nu}} \cdot \boldsymbol{\beta}
+2 ( \boldsymbol{\hat{\nu}}\cdot\boldsymbol{\hat{q}} )
( \boldsymbol{\beta} \cdot \boldsymbol{\hat{q}} )
\Big)
\vert \langle J_f \Vert \mathcal{\hat{L}}_J (q) \Vert J_i\rangle \vert^2
\nonumber\\
&+
\Big(1+\boldsymbol{\hat{\nu}} \cdot \boldsymbol{\beta} \Big)
\vert \langle J_f \Vert \mathcal{\hat{M}}_J (q) \Vert J_i\rangle \vert^2-
\boldsymbol{\hat{q}}\cdot ( \boldsymbol{\hat{\nu}}+\boldsymbol{\beta})
\nonumber\\
&
\times 2 \mathcal{R}e \bigg(
\langle J_f \Vert \mathcal{\hat{L}}_J (q) \Vert J_i\rangle
\langle J_f \Vert \mathcal{\hat{M}}_J (q) \Vert J_i\rangle^*
\bigg)
\bigg],
\label{eq:M^0}
\end{align}
while the second is the electrostatically corrected multipole expansion:
\begin{align}
\mathcal{M}^c
&=
\pm 4E_e\!\int_{0}^{\infty}\! dk_\gamma\,k_\gamma^{2} \dfrac{1}{(2\pi)^3} \int d\Omega_{\gamma} 
\mathcal{R}e\Bigg\{
\Theta(k,k_\gamma)
\nonumber\\
&\times
\Bigg[
\sum_{J\ge1}\bigg[
\Big(1-(\boldsymbol{\hat{\nu}}\cdot\boldsymbol{\hat{q}}')
(\boldsymbol{\beta}\cdot\boldsymbol{\hat{q}}')\Big)
\nonumber\\
&\times
\Big(
\langle J_f\Vert \hat{\mathcal{T}}_{J}^{\mathrm{mag}}(q)\Vert J_i\rangle
\langle J_f\Vert \hat{\mathcal{T}}_{J}^{\mathrm{mag}}(q')\Vert J_i\rangle^{*}
\nonumber\\
&+
\langle J_f\Vert \hat{\mathcal{T}}_{J}^{\mathrm{el}}(q)\Vert J_i\rangle
\langle J_f\Vert \hat{\mathcal{T}}_{J}^{\mathrm{el}}(q')\Vert J_i\rangle^{*}
\Big)
\nonumber\\
&\pm
\Big(\boldsymbol{\hat{q}}'\cdot(\boldsymbol{\hat{\nu}}-\boldsymbol{\beta})\Big)
\Big(
\langle J_f\Vert \hat{\mathcal{T}}_{J}^{\mathrm{mag}}(q)\Vert J_i\rangle
\langle J_f\Vert \hat{\mathcal{T}}_{J}^{\mathrm{el}}(q')\Vert J_i\rangle^{*}
\nonumber\\
&+
\langle J_f\Vert \hat{\mathcal{T}}_{J}^{\mathrm{el}}(q)\Vert J_i\rangle
\langle J_f\Vert \hat{\mathcal{T}}_{J}^{\mathrm{mag}}(q')\Vert J_i\rangle^{*}
\Big)
\bigg]
\nonumber\\
&+
\sum_{J\ge0}\bigg[
\Big(
1-\boldsymbol{\hat{\nu}}\cdot\boldsymbol{\beta}
+2(\boldsymbol{\hat{\nu}}\cdot\boldsymbol{\hat{q}}')
(\boldsymbol{\beta}\cdot\boldsymbol{\hat{q}}')
\Big)
\nonumber\\
&\times
\langle J_f\Vert \hat{\mathcal{L}}_{J}(q)\Vert J_i\rangle
\langle J_f\Vert \hat{\mathcal{L}}_{J}(q')\Vert J_i\rangle^{*}
\nonumber\\
&+
\Big(1+\boldsymbol{\hat{\nu}}\cdot\boldsymbol{\beta}\Big)
\langle J_f\Vert \hat{\mathcal{M}}_{J}(q)\Vert J_i\rangle
\langle J_f\Vert \hat{\mathcal{M}}_{J}(q')\Vert J_i\rangle^{*}
\nonumber\\
&-
\Big(\boldsymbol{\hat{q}}'\cdot(\boldsymbol{\hat{\nu}}+\boldsymbol{\beta})\Big)
\langle J_f\Vert \hat{\mathcal{L}}_{J}(q)\Vert J_i\rangle
\langle J_f\Vert \hat{\mathcal{M}}_{J}(q')\Vert J_i\rangle^{*} 
\nonumber\\
&-
\Big(\boldsymbol{\hat{q}}'\cdot(\boldsymbol{\hat{\nu}}+\boldsymbol{\beta})\Big)
\langle J_f\Vert \hat{\mathcal{M}}_{J}(q)\Vert J_i\rangle
\langle J_f\Vert \hat{\mathcal{L}}_{J}(q')\Vert J_i\rangle^{*}
\bigg]
\Bigg]
\Bigg\}.
\label{eq:electrostatic correction}
\end{align}
We note that this Coulomb correction comprises three distinct contributions: the inclusion of the nuclear charge form factor, the shift in the momentum transfer that modifies the leptonic traces, and a similar shift in the momentum transfer that enters the multipole operators.

\section{\textbf{The $\beta$-decay rate incorporating matrix elements, multipole operators, and Coulomb-interaction corrections}}
\label{sec_3}

\begin{table*}[t] 
\raggedright
\caption{Leptonic traces with Coulomb-interaction corrections for $\beta^{\mp}$ decay to $\mathcal{O}(k_\gamma)$, where $\boldsymbol{\beta} = \dfrac{\boldsymbol{k}}{E_e}$, 
$\hat{\boldsymbol{\nu}} = \dfrac{\boldsymbol{\nu}}{E_\nu}$, 
and $\hat{\boldsymbol{q}} = \dfrac{\boldsymbol{q}}{q}$.}
\label{tab:lepton-traces-kp}
\centering
\renewcommand{\arraystretch}{1.8}

\resizebox{\textwidth}{!}{
\begin{tabular}{|c|c|}
\hline
\textbf{Summand} & \textbf{Coulomb interaction correction to $\mathcal{O}(k_\gamma)$} \\
\hline

$\dfrac{1}{2}\Big( \boldsymbol{l}\cdot\boldsymbol{L}^{c*}-l_3L_3^{c*}\Big)$
&
$2E_e\Big[1-(\hat{\boldsymbol{\nu}}\cdot\hat{\boldsymbol{q}})
(\boldsymbol{\beta}\cdot\hat{\boldsymbol{q}})
+\frac{1}{q}\Big((\hat{\boldsymbol{\nu}}\cdot\hat{\boldsymbol{q}})
(\boldsymbol{\beta}\cdot\boldsymbol{k}_\gamma)
+(\boldsymbol{\beta}\cdot\hat{\boldsymbol{q}})
(\hat{\boldsymbol{\nu}}\cdot\boldsymbol{k}_\gamma)
-2(\hat{\boldsymbol{\nu}}\cdot\hat{\boldsymbol{q}})
(\boldsymbol{\beta}\cdot\hat{\boldsymbol{q}})
(\boldsymbol{k}_\gamma\cdot\hat{\boldsymbol{q}})\Big)\Big]$
\\
\hline

$l_0L_0^{c*}$ &
$2E_e\big[1+\hat{\boldsymbol{\nu}}\cdot\boldsymbol{\beta}\big]$
\\
\hline

$l_3L_3^{c*}$ &
$2E_e\Big[1-\hat{\boldsymbol{\nu}}\cdot\boldsymbol{\beta}
+2(\hat{\boldsymbol{\nu}}\cdot\hat{\boldsymbol{q}})
(\boldsymbol{\beta}\cdot\hat{\boldsymbol{q}})
-\frac{2}{q}\Big((\hat{\boldsymbol{\nu}}\cdot\hat{\boldsymbol{q}})
(\boldsymbol{\beta}\cdot\boldsymbol{k}_\gamma)
+(\boldsymbol{\beta}\cdot\hat{\boldsymbol{q}})
(\hat{\boldsymbol{\nu}}\cdot\boldsymbol{k}_\gamma)
-2(\hat{\boldsymbol{\nu}}\cdot\hat{\boldsymbol{q}})
(\boldsymbol{\beta}\cdot\hat{\boldsymbol{q}})
(\boldsymbol{k}_\gamma\cdot\hat{\boldsymbol{q}})\Big)\Big]$
\\
\hline

$-\Big(l_3L_0^{c*}\Big)$ &
$2E_e\Big[-\hat{\boldsymbol{q}}\cdot(\hat{\boldsymbol{\nu}}+\boldsymbol{\beta})
+\frac{1}{q}(\boldsymbol{k}_\gamma\cdot\hat{\boldsymbol{\nu}}+\boldsymbol{k}_\gamma\cdot\boldsymbol{\beta})
-\frac{\boldsymbol{k}_\gamma\cdot\hat{\boldsymbol{q}}}{q}
(\hat{\boldsymbol{q}}\cdot\hat{\boldsymbol{\nu}}+\hat{\boldsymbol{q}}\cdot\boldsymbol{\beta})\Big]$
\\
\hline

$-\dfrac{i}{2}(\boldsymbol{l}\times\boldsymbol{L}^{c*})_3$ &
$ \mp 2E_e\Big[-\hat{\boldsymbol{q}}\cdot(\hat{\boldsymbol{\nu}}-\boldsymbol{\beta})
+\frac{1}{q}(\boldsymbol{k}_\gamma\cdot\hat{\boldsymbol{\nu}}-\boldsymbol{k}_\gamma\cdot\boldsymbol{\beta})
-\frac{\boldsymbol{k}_\gamma\cdot\hat{\boldsymbol{q}}}{q}
(\hat{\boldsymbol{q}}\cdot\hat{\boldsymbol{\nu}}-\hat{\boldsymbol{q}}\cdot\boldsymbol{\beta})\bigg]$
\\
\hline

$-\Big(l_0L_3^{c*}\Big)$ &
$2E_e\Big[-\hat{\boldsymbol{q}}\cdot(\hat{\boldsymbol{\nu}}+\boldsymbol{\beta})
+\frac{1}{q}(\boldsymbol{k}_\gamma\cdot\hat{\boldsymbol{\nu}}+\boldsymbol{k}_\gamma\cdot\boldsymbol{\beta})
-\frac{\boldsymbol{k}_\gamma\cdot\hat{\boldsymbol{q}}}{q}
(\hat{\boldsymbol{q}}\cdot\hat{\boldsymbol{\nu}}+\hat{\boldsymbol{q}}\cdot\boldsymbol{\beta})\Big]$
\\
\hline
\end{tabular}
}
\end{table*}

We first consider the expansion of the Coulomb-shifted momentum $\boldsymbol{q}'$ and its direction $\hat{\boldsymbol{q}}'$. To first order in $k_\gamma/q$, one gets
\begin{equation}
\label{eq:q'}
\hat{\boldsymbol{q}}'
\simeq
\hat{\boldsymbol{q}}
-
\frac{\boldsymbol{k}_\gamma}{q}
+
\frac{\boldsymbol{k}_\gamma \cdot \hat{\boldsymbol{q}}}{q}\,
\hat{\boldsymbol{q}} .
\end{equation}
This allows all angular structures involving $\hat{\boldsymbol{q}}'$ to be expanded systematically to this order, with the corresponding leptonic trace contributions summarized in Table~\ref{tab:lepton-traces-kp}.

Noting the three distinct sources of corrections entering the Coulomb part of the $\beta$ decay amplitude, and restricting attention to the leading correction, we can decompose the Coulomb contribution as:
\begin{equation}
    \label{eq:M^c}
    \mathcal{M}^{c}
    =
    \mathcal{M}^{c}_{\mathrm{FF}}
    +
    \mathcal{M}^{c}_{\mathrm{Tr}}
    +
    \mathcal{M}^{c}_{\mathrm{Op}},
\end{equation}
where the three terms correspond to the correction induced by the nuclear charge form factor, the modification of the leptonic traces, and the shift in the momentum transfer entering the multipole operators, respectively. In each case, only the corresponding contribution is expanded to first order in $k_\gamma$, while all other contributions are kept at zeroth order.

Noting that in the zeroth order $\hat{\boldsymbol q}' \rightarrow \hat{\boldsymbol q}$, yielding the standard Walecka lepton traces and the standard multipole operators (as they appear in Eq.~\eqref{eq:M^0}), we obtain from Eq.~\eqref{eq:electrostatic correction} the first Coulomb correction, associated with the nuclear charge form factor:
\begin{align}
        &\mathcal{M}^{c}_{\mathrm{FF}} = \pm 4 E_e \int_{0}^{\infty}  dk_{\gamma} k_{\gamma}^{2} \dfrac{1}{(2\pi)^3} \int d\Omega_\gamma
\mathcal{R}e \Theta(k,k_\gamma) \nonumber\\
&\times \Bigg\{ 
\sum_{J\ge1} \bigg[
 \Big(  \big(1 
- (\hat{\boldsymbol{\nu}} \cdot \hat{\boldsymbol{q}})
  (\boldsymbol{\beta} \cdot \hat{\boldsymbol{q}}) \big) \Big)\nonumber\\
 &\times \Big(\big|\langle J_f \Vert \mathcal{\hat{T}}^{mag}_{J}(q) \Vert J_i \rangle\big|^{2} +  
\big| \big\langle J_{f}\big\Vert \hat{\mathcal{T}}_{J}^{\mathrm{el}}(q)
\big\Vert J_{i}\big\rangle \big|^2 \Big)  \nonumber\\
&  \pm  \hat{\boldsymbol{q}} \cdot (\hat{\boldsymbol{\nu}} - \boldsymbol{\beta})  2 \mathcal{R}e 
\big\langle J_{f}\big\Vert \hat{\mathcal{T}}_{J}^{\mathrm{mag}}(q)
\big\Vert J_{i}\big\rangle \big\langle J_{f}\big\Vert \hat{\mathcal{T}}_{J}^{\mathrm{el}}(q)
\big\Vert J_{i}\big\rangle^{*} \bigg] \nonumber\\
& +
\sum_{J\ge0}
\bigg[ \Big( 1 - \hat{\boldsymbol{\nu}} \cdot \boldsymbol{\beta}
+ 2(\hat{\boldsymbol{\nu}} \cdot \hat{\boldsymbol{q}})
    (\boldsymbol{\beta} \cdot \hat{\boldsymbol{q}}) \Big)
\big| \big\langle J_{f}\big\Vert \hat{\mathcal{L}}_{J}(q)
\big\Vert J_{i}\big\rangle \big|^2 \nonumber\\
& + \big(1+\boldsymbol{\hat{\nu}} \cdot \boldsymbol{\beta} \big)
\big|\big\langle J_{f}\big\Vert \hat{\mathcal{M}}_{J}(q)
\big\Vert J_{i}\big\rangle \big|^2 \nonumber\\
& -  \hat{\boldsymbol{q}} \cdot (\hat{\boldsymbol{\nu}} + \boldsymbol{\beta}) \big) 2 \mathcal{R}e 
\big\langle J_{f}\big\Vert \hat{\mathcal{L}}_{J}(q)
\big\Vert J_{i}\big\rangle \big\langle J_{f}\big\Vert \hat{\mathcal{M}}_{J}(q)
\big\Vert J_{i}\big\rangle^{*}  \bigg]  \Bigg \rbrace \nonumber\\
& + \mathcal{O}\big(k_\gamma^2\big)
\label{eq:FF_1}
    \end{align}
where $\mathcal{O}(k_\gamma^2)$ collectively denotes higher-order terms in the exchanged-photon momentum expansion.

Evaluating the angular and exchanged-photon momentum integrals,
including the nuclear charge form factor of Eq.~\eqref{eq:FF} and the
functions defined in Eq.~\eqref{eq:Theta}, we obtain:
\begin{multline}
    \pm 4 E_e \int_{0}^{\infty}  dk_{\gamma} k_{\gamma}^{2} \dfrac{1}{(2\pi)^3} \int d\Omega_\gamma
    \mathcal{R}e \Theta(k,k_\gamma)\\
    = \pm \pi \alpha Z \dfrac{E_e}{k} \mp 4 \alpha Z a E_e .
\end{multline}
Identifying the leading-order expansion of the Fermi function in the parameter $\alpha Z/k$, the point-Coulomb term can be written as
\begin{equation}
    \pm
    \pi\alpha Z\,\frac{E_e}{k}
    =
    F^{\mp}(Z,E_e)-1
    +
    \mathcal{O}\!\left((\alpha Z/k)^2\right).
\end{equation}
Since the point-charge Fermi function is a function of a different parameter $\alpha Z/k$, independent of nuclear structure, we can take it to all orders in this parameter. 
Substituting the integrated result into Eq.~\eqref{eq:FF_1} and using the
definition of $\mathcal{M}^{0}$, we obtain the leading nuclear structure dependent Coulomb
contribution $\mathcal{M}^{c}_{\mathrm{FF}}$ generated by the nuclear charge
form factor,
\begin{equation}
    \mathcal{M}^{c}_{\mathrm{FF}} =  \Big[ F^{\mp}(Z,E_e)-1
    \mp
    4\alpha Z a E_e \Big] \mathcal{M}^{0}.
    \label{eq:M^c_FF}
\end{equation}
Equation~\eqref{eq:M^c_FF} contains two distinct contributions.
The first corresponds to the leading-order expansion of the standard
point-charge Fermi function, while the second represents a finite-size
correction of order $\alpha Z E_e R$ (since $a=R/\sqrt{6}$).
In other words, the nuclear-independent Coulomb effects are absorbed into the exact
Fermi function $F(Z,E_e)$, whereas the nuclear-structure contribution
is treated perturbatively in the small parameter $\varepsilon_{EC}$.

Returning to the electrostatically corrected expression, Eq.~\eqref{eq:electrostatic correction}, we evaluate the Coulomb-induced corrections to the Walecka lepton traces within the momentum-shift expansion, employing the results appearing in Table~\ref{tab:lepton-traces-kp}, and obtain
\begin{align}
    &\mathcal{M}^{c}_{\mathrm{Tr}} = \pm 4 E_e \int_{0}^{\infty}  dk_{\gamma} k_{\gamma}^{2} \dfrac{1}{(2\pi)^3} \int d\Omega_\gamma
    \mathcal{R}e \Theta(k,k_\gamma) \nonumber\\
    &\times \dfrac{1}{q} \Bigg\{ 
    \sum_{J\ge1} \bigg[
     \Big( 
    (\hat{\boldsymbol{\nu}} \cdot \hat{\boldsymbol{q}})
      (\boldsymbol{\beta} \cdot \boldsymbol{k}_\gamma) +
    (\boldsymbol{\beta} \cdot \hat{\boldsymbol{q}})
      (\hat{\boldsymbol{\nu}} \cdot \boldsymbol{k}_\gamma) \nonumber\\
      & - 2 ( \hat{\boldsymbol{\nu}} \cdot \hat{\boldsymbol{q}})
      (\boldsymbol{\beta} \cdot \hat{\boldsymbol{q}})
      (\boldsymbol{k}_\gamma \cdot \hat{\boldsymbol{q}}) \Big) \Big(\big|\langle J_f \Vert \mathcal{\hat{T}}^{mag}_{J}(q) \Vert J_i \rangle\big|^{2} \nonumber\\
      & +  
    \big| \big\langle J_{f}\big\Vert \hat{\mathcal{T}}_{J}^{\mathrm{el}}(q)
    \big\Vert J_{i}\big\rangle \big|^2 \Big) \nonumber \\
    &  \pm  \Big( 
    - \boldsymbol{k}_\gamma \cdot \hat{\boldsymbol{\nu}}
          + \boldsymbol{k}_\gamma \cdot \boldsymbol{\beta} 
    + (\boldsymbol{k}_\gamma \cdot \hat{\boldsymbol{q}})
      (\hat{\boldsymbol{q}} \cdot \hat{\boldsymbol{\nu}}
          - \hat{\boldsymbol{q}} \cdot \boldsymbol{\beta}) \Big) \nonumber\\
          & \times 2 \mathcal{R}e 
    \big\langle J_{f}\big\Vert \hat{\mathcal{T}}_{J}^{\mathrm{mag}}(q)
    \big\Vert J_{i}\big\rangle \big\langle J_{f}\big\Vert \hat{\mathcal{T}}_{J}^{\mathrm{el}}(q)
    \big\Vert J_{i}\big\rangle^{*}  \bigg] \nonumber\\
    & +
    \sum_{J\ge0}
    \bigg[ - 2 \Big( 
    (\hat{\boldsymbol{\nu}} \cdot \hat{\boldsymbol{q}})
       (\boldsymbol{\beta} \cdot \boldsymbol{k}_\gamma) +
    (\boldsymbol{\beta} \cdot\hat{\boldsymbol{q}})
       (\hat{\boldsymbol{\nu}} \cdot \boldsymbol{k}_\gamma) \nonumber\\
       & -
    2(\hat{\boldsymbol{\nu}} \cdot \hat{\boldsymbol{q}})
      (\boldsymbol{\beta} \cdot \hat{\boldsymbol{q}})
      (\boldsymbol{k}_\gamma \cdot \hat{\boldsymbol{q}})
     \Big) \big| \big\langle J_{f}\big\Vert \hat{\mathcal{L}}_{J}(q)
    \big\Vert J_{i}\big\rangle \big|^2 \nonumber\\
    & +
    \Big( \boldsymbol{k}_\gamma \cdot \hat{\boldsymbol{\nu}}
          + \boldsymbol{k}_\gamma  \cdot \boldsymbol{\beta} 
    - (\boldsymbol{k}_\gamma \cdot \hat{\boldsymbol{q}})
      (\hat{\boldsymbol{q}} \cdot \hat{\boldsymbol{\nu}}
          + \hat{\boldsymbol{q}} \cdot \boldsymbol{\beta} ) \Big) \nonumber\\
          & \times 2 \mathcal{R}e 
    \big\langle J_{f}\big\Vert \hat{\mathcal{L}}_{J}(q)
    \big\Vert J_{i}\big\rangle \big\langle J_{f}\big\Vert \hat{\mathcal{M}}_{J}(q)
    \big\Vert J_{i}\big\rangle^{*}  \bigg] \Bigg\} + \mathcal{O}\big(k_\gamma^2\big)
    \label{eq:M^c_Tr1}
\end{align}
We note that the leptonic structures arising from the momentum-shift
expansion, listed in Table~\ref{tab:lepton-traces-kp}, generate the
angular integrals appearing in Eq.~\eqref{eq:M^c_Tr1}, i.e.,
$(\boldsymbol{k}_\gamma \cdot \hat{\boldsymbol{q}})$,
$(\boldsymbol{\beta} \cdot \boldsymbol{k}_\gamma)$,
and $(\hat{\boldsymbol{\nu}} \cdot \boldsymbol{k}_\gamma)$.
Evaluating the angular and exchanged-photon momentum integrals,
\begin{align}
    &\pm
    4E_e
    \int_{0}^{\infty} dk_\gamma\,k_\gamma^2
    \frac{1}{(2\pi)^3}
    \int d\Omega_\gamma\,
    \mathcal{R}e\,\Theta(k,k_\gamma)
    \nonumber\\
    &\qquad\times
    \Big\{
    (\boldsymbol{k}_\gamma \cdot \hat{\boldsymbol q}),
    \,
    (\boldsymbol{\beta} \cdot \boldsymbol{k}_\gamma),
    \,
    (\hat{\boldsymbol\nu} \cdot \boldsymbol{k}_\gamma)
    \Big\} \nonumber\\
    & \qquad= \mp \alpha Z a k E_e \dfrac{16}{3} \Big\{
    (\hat{\boldsymbol{k}} \cdot \hat{\boldsymbol q}),
    \,
    (\boldsymbol{\beta} \cdot \hat{\boldsymbol{k}}),
    \,
    (\hat{\boldsymbol\nu} \cdot \hat{\boldsymbol{k}})
    \Big\},
\end{align}
and expressing
$\hat{\boldsymbol{k}}=(E_e/k)\boldsymbol{\beta}$, yields the second
Coulomb correction term, associated with the leptonic-trace momentum-shift expansion,
\begin{align}
    &\mathcal{M}^{c}_{\mathrm{Tr}} = \mp \alpha Z a\dfrac{ E_e^2}{q}  \dfrac{16}{3} \Bigg\{ 
    \sum_{J\ge1} \bigg[
     \Big(
    (\hat{\boldsymbol{\nu}} \cdot \hat{\boldsymbol{q}})
      \beta^2 +
    (\boldsymbol{\beta} \cdot \hat{\boldsymbol{q}})
      (\hat{\boldsymbol{\nu}} \cdot \boldsymbol{\beta}) \nonumber \\
      & - 2 ( \hat{\boldsymbol{\nu}} \cdot \hat{\boldsymbol{q}})
      (\boldsymbol{\beta} \cdot \hat{\boldsymbol{q}})^2
     \Big) \Big(\big|\langle J_f \Vert \mathcal{\hat{T}}^{mag}_{J}(q) \Vert J_i \rangle\big|^{2} \nonumber \\
     & +  
    \big| \big\langle J_{f}\big\Vert \hat{\mathcal{T}}_{J}^{\mathrm{el}}(q)
    \big\Vert J_{i}\big\rangle \big|^2 \Big) \nonumber \\
    &\pm  \Big(\boldsymbol{\beta}^2 -\boldsymbol{\beta} \cdot \hat{\boldsymbol\nu}
    +
    (\boldsymbol{\beta} \cdot \hat{\boldsymbol q})
    (\hat{\boldsymbol q} \cdot \hat{\boldsymbol\nu}
    -\hat{\boldsymbol q} \cdot \boldsymbol\beta) \Big) \nonumber \\
    &\times
      2 \mathcal{R}e \Big(
    \big\langle J_{f}\big\Vert \hat{\mathcal{T}}_{J}^{\mathrm{mag}}(q)
    \big\Vert J_{i}\big\rangle \big\langle J_{f}\big\Vert \hat{\mathcal{T}}_{J}^{\mathrm{el}}(q)
    \big\Vert J_{i}\big\rangle^{*} \Big) \bigg] \nonumber \\
    & +
    \sum_{J\ge0}
    \bigg[  - 2
    \Big(
    (\hat{\boldsymbol{\nu}} \cdot \hat{\boldsymbol{q}})
       \beta^2 +
    (\boldsymbol{\beta} \cdot\hat{\boldsymbol{q}})
       (\hat{\boldsymbol{\nu}} \cdot \boldsymbol{\beta})  -
    2(\hat{\boldsymbol{\nu}} \cdot \hat{\boldsymbol{q}})
      (\boldsymbol{\beta} \cdot \hat{\boldsymbol{q}})^2
    \Big)\nonumber \\
    &\times
    \big| \big\langle J_{f}\big\Vert \hat{\mathcal{L}}_{J}(q)
    \big\Vert J_{i}\big\rangle \big|^2  + \Big(\boldsymbol{\beta}^2 + \boldsymbol{\beta} \cdot \hat{\boldsymbol\nu}
    -
    (\boldsymbol{\beta} \cdot \hat{\boldsymbol q})
    (\hat{\boldsymbol q} \cdot \hat{\boldsymbol\nu}
    -\hat{\boldsymbol q} \cdot \boldsymbol\beta) \Big)\nonumber \\
    &\times   2 \mathcal{R}e \Big(
    \big\langle J_{f}\big\Vert \hat{\mathcal{L}}_{J}(q)
    \big\Vert J_{i}\big\rangle \big\langle J_{f}\big\Vert \hat{\mathcal{M}}_{J}(q)
    \big\Vert J_{i}\big\rangle^{*} \Big) \bigg] \Bigg\}.
    \label{eq:M^c_Tr}
\end{align}
The perturbative contribution $\mathcal{M}^{c}_{\mathrm{Tr}}$ therefore exhibits
an explicit $\mathcal{O}(\varepsilon_{EC})$ dependence, together with
modified angular correlations.
Notably absent is a contribution proportional to
$\big| \big\langle J_{f}\big\Vert \hat{\mathcal{M}}_{J}(q) \big\Vert J_{i}\big\rangle \big|^2$, which governs, for example, the pure Fermi $\beta$-decay transition. This is because the corresponding leptonic trace is independent of the momentum transfer $q$, so no leptonic-trace correction arises for this term.

We turn to evaluate the third and final Coulomb correction,
$\mathcal{M}^c_{\mathrm{Op}}$, arising from the Coulomb-induced shift of the
momentum transfer in the multipole operators.
The corresponding shift of the multipole operators may be expanded as
\begin{align}
    & \langle J_f\Vert\{\mathcal{M}_{J},\,\mathcal{L}_{J},\,\mathcal{T}^{el}_{J},\,\mathcal{T}^{mag}_{J}\}(q')\Vert J_i\rangle \nonumber \\
    & = \langle J_f\Vert\{\mathcal{M}_{J},\,\mathcal{L}_{J},\,\mathcal{T}^{el}_{J},\,\mathcal{T}^{mag}_{J}\}(q)\Vert J_i\rangle \nonumber \\
    & - (\boldsymbol{k}_\gamma \cdot \hat{\boldsymbol{q}})\,
    \frac{d}{d q}\langle J_f\Vert\{\mathcal{M}_{J},\,\mathcal{L}_{J},\,\mathcal{T}^{el}_{J},\,\mathcal{T}^{mag}_{J}\}(q)\Vert J_i\rangle
    + \mathcal{O}(k_\gamma^2).
    \label{eq:Multipole Exp}
\end{align}
The  derivatives of the multipole operators are evaluated using a known relation for the spherical Bessel function:
\begin{equation}
\begin{aligned}
\label{eq:multipole shift}
&
\frac{d}{dq}
\left\langle J_f\Big\Vert
\left\{
\mathcal{M}_{J},
\mathcal{L}_{J},
\mathcal{T}^{el}_{J},
\mathcal{T}^{mag}_{J}
\right\}(q)
\Big\Vert J_i
\right\rangle
\\
&=
\left\{
\begin{aligned}
&
\frac{J}{q}
\left\langle J_f\Big\Vert
\left\{
\mathcal{M}_{J},
\mathcal{T}^{mag}_{J}
\right\}(q)
\Big\Vert J_i
\right\rangle
\\[2mm]
&
\frac{J-1}{q}
\left\langle J_f\Big\Vert
\left\{
\mathcal{L}_{J},
\mathcal{T}^{el}_{J}
\right\}(q)
\Big\Vert J_i
\right\rangle
\end{aligned}
\right.
\\
&\qquad\times
\Big[
1+\mathcal{O}\!\left((qR)^2\right)
\Big]
\end{aligned}
\end{equation}

Applying the multipole shift and Eq.~\eqref{eq:multipole shift} to the relevant terms in Eq.~\eqref{eq:electrostatic correction}, together with the standard leptonic traces, 
we obtain the contribution to $\mathcal{M}^{c}_{\mathrm{Op}}$ arising from the Coulomb-induced shift of the momentum transfer in the multipole operators,
\begin{align}
        &\mathcal{M}^{c}_{\mathrm{Op}} = \mp 4 E_e\!\int_{0}^{\infty}  dk_{\gamma}\,k_{\gamma}^{2} \dfrac{1}{(2\pi)^3}\int d\Omega_{\gamma} \mathcal{R}e \Theta(k,k_\gamma)(\boldsymbol{k_\gamma} \cdot \hat{\boldsymbol{q}}) \nonumber\\
&\times        
 \Bigg\{
\sum_{J\ge1} \bigg[ \Big(1-\big( \boldsymbol{\hat{\nu}} \cdot \boldsymbol{\hat{q}}\big)\big(\boldsymbol{\beta} \cdot \boldsymbol{\hat{q}}\big) \Big) \nonumber\\
&\times \bigg( \dfrac{J}{q} \big| \big\langle J_{f}\big\Vert \hat{\mathcal{T}}_{J}^{\mathrm{mag}}(q)
\big\Vert J_{i}\big\rangle \big|^2 + \dfrac{J-1}{q} \big| \big\langle J_{f}\big\Vert \hat{\mathcal{T}}_{J}^{\mathrm{el}}(q)
\big\Vert J_{i}\big\rangle \big|^2 \bigg) \nonumber\\
& \pm   \boldsymbol{\hat{q}}\cdot \big(\boldsymbol{\hat{\nu}}-\boldsymbol{\beta}\big)  \nonumber \\
&\times \bigg( \frac{J-1}{q} \big\langle J_{f}\big\Vert \hat{\mathcal{T}}_{J}^{\mathrm{mag}}(q)
\big\Vert J_{i}\big\rangle
\big\langle J_{f}\big\Vert \hat{\mathcal{T}}_{J}^{\mathrm{el}}(q)
\big\Vert J_{i}\big\rangle^{*}  \nonumber\\
& + \frac{J}{q} \big\langle J_{f}\big\Vert \hat{\mathcal{T}}_{J}^{\mathrm{el}}(q)
\big\Vert J_{i}\big\rangle
\big\langle J_{f}\big\Vert \hat{\mathcal{T}}_{J}^{\mathrm{mag}}(q)
\big\Vert J_{i}\big\rangle^{*} \bigg)
\bigg] \nonumber\\
& +
\sum_{J\ge0}
\bigg[ \Big( 1-\boldsymbol{\hat{\nu}} \cdot \boldsymbol{\beta}+2 \big(\boldsymbol{\hat{\nu}}\cdot\boldsymbol{\hat{q}}\big)\big(\boldsymbol{\beta} \cdot \boldsymbol{\hat{q}}\big) \Big) \nonumber\\
&\times
\dfrac{J-1}{q}
 \big| \big\langle J_{f}\big\Vert \hat{\mathcal{L}}_{J}(q)
\big\Vert J_{i}\big\rangle \big|^2\nonumber \\
& + 
\big(1+\boldsymbol{\hat{\nu}} \cdot \boldsymbol{\beta} \big) \dfrac{J}{q} \big|  \big\langle J_{f}\big\Vert \hat{\mathcal{M}}_{J}(q)
\big\Vert J_{i}\big\rangle \big|^2 -  \boldsymbol{\hat{q}}\cdot \big(\boldsymbol{\hat{\nu}}+\boldsymbol{\beta}\big) \nonumber \\
& \times 
\bigg( \frac{J}{q}
\big\langle J_{f}\big\Vert \hat{\mathcal{L}}_{J}(q)
\big\Vert J_{i}\big\rangle
\big\langle J_{f}\big\Vert \hat{\mathcal{M}}_{J}(q)
\big\Vert J_{i}\big\rangle^{*} \nonumber\\
& + 
\frac{J-1}{q} \big\langle J_{f}\big\Vert \hat{\mathcal{M}}_{J}(q)
\big\Vert J_{i}\big\rangle
\big\langle J_{f}\big\Vert \hat{\mathcal{L}}_{J}(q)
\big\Vert J_{i}\big\rangle^{*} \bigg) \bigg]\Bigg\} \nonumber\\
& + \mathcal{O}(k_\gamma^2) + \mathcal{O}\big((q R)^2\big).
\label{eq:Op_1}
\end{align}
\\
Performing the angular and exchanged-photon momentum integrations yields
\begin{multline}
    \mp 4 E_e\!\int_{0}^{\infty}  dk_{\gamma}\,k_{\gamma}^{2} \dfrac{1}{(2\pi)^3}\int d\Omega_{\gamma} \Theta(k,k_\gamma)(\boldsymbol{k_\gamma} \cdot \hat{\boldsymbol{q}}) \\
    = \pm \alpha Z a E_e^2 \dfrac{16}{3} 
\big(\boldsymbol{\beta} \cdot \hat{\boldsymbol{q}}\big).
\end{multline}
Substituting the result of the integral into Eq.~\eqref{eq:Op_1} and simplifying, we obtain
    \begin{align}
        &\mathcal{M}^{c}_{\mathrm{Op}} = \pm \alpha Z a  E_e^2 \dfrac{1}{q}\dfrac{16}{3} 
\big(\boldsymbol{\beta} \cdot \hat{\boldsymbol{q}}\big) \nonumber\\
&\times        
 \Bigg\{
\sum_{J\ge1} \bigg[ \Big(1-\big( \boldsymbol{\hat{\nu}} \cdot \boldsymbol{\hat{q}}\big)\big(\boldsymbol{\beta} \cdot \boldsymbol{\hat{q}}\big) \Big) \nonumber\\
&\times \Big( J \big| \big\langle J_{f}\big\Vert \hat{\mathcal{T}}_{J}^{\mathrm{mag}}(q)
\big\Vert J_{i}\big\rangle \big|^2 + (J-1) \big| \big\langle J_{f}\big\Vert \hat{\mathcal{T}}_{J}^{\mathrm{el}}(q)
\big\Vert J_{i}\big\rangle \big|^2 \Big)\nonumber \\
& \pm  \boldsymbol{\hat{q}}\cdot \big(\boldsymbol{\hat{\nu}}-\boldsymbol{\beta}\big)   \nonumber\\
&\times  (2J-1) \mathcal{R}e\big\langle J_{f}\big\Vert \hat{\mathcal{T}}_{J}^{\mathrm{mag}}(q)
\big\Vert J_{i}\big\rangle
\big\langle J_{f}\big\Vert \hat{\mathcal{T}}_{J}^{\mathrm{el}}(q)
\big\Vert J_{i}\big\rangle^{*} \bigg]\nonumber \\
& +
\sum_{J\ge0}
\bigg[ \Big( 1-\boldsymbol{\hat{\nu}} \cdot \boldsymbol{\beta}+2 \big(\boldsymbol{\hat{\nu}}\cdot\boldsymbol{\hat{q}}\big)\big(\boldsymbol{\beta} \cdot \boldsymbol{\hat{q}}\big) \Big) \nonumber\\
&\times
(J-1)
 \big| \big\langle J_{f}\big\Vert \hat{\mathcal{L}}_{J}(q)
\big\Vert J_{i}\big\rangle \big|^2 \nonumber\\
& + 
\big(1+\boldsymbol{\hat{\nu}} \cdot \boldsymbol{\beta} \big) J \big|  \big\langle J_{f}\big\Vert \hat{\mathcal{M}}_{J}(q)
\big\Vert J_{i}\big\rangle \big|^2 - \boldsymbol{\hat{q}}\cdot \big(\boldsymbol{\hat{\nu}}+\boldsymbol{\beta}\big)  \nonumber\\
& \times 
 (2J - 1) \mathcal{R}e
\big\langle J_{f}\big\Vert \hat{\mathcal{L}}_{J}(q)
\big\Vert J_{i}\big\rangle
\big\langle J_{f}\big\Vert \hat{\mathcal{M}}_{J}(q)
\big\Vert J_{i}\big\rangle^{*} \bigg] \Bigg\}\nonumber \\
& + \mathcal{O}(k_\gamma^2) + \mathcal{O}\big((q R)^2\big) 
\label{eq:Op}
    \end{align}

The third contribution, $\mathcal{M}^{c}_{\mathrm{Op}}$, exhibits the same angular structure as the standard Walecka term $\mathcal{M}^{0}$, but is additionally suppressed by the finite-size Coulomb factor $\varepsilon_{EC}$. Unlike $\mathcal{M}^{0}$, the multipole rank $J$, which characterizes the total angular-momentum transfer of the transition, appears explicitly as a weighting factor multiplying the multipole operators. Consequently, $\mathcal{M}^{c}_{\mathrm{Op}}$ constitutes a transition-dependent correction whose magnitude depends explicitly on the contributing multipole rank and therefore on the specific $\beta$-decay channel.

We now proceed to the final stage in the formulation of the $\beta$-decay rate for general transitions. Substituting the Coulomb contributions of Eq.~\eqref{eq:M^c}, with the explicit connection of the first contribution to the known Fermi function (Eq.~\eqref{eq:M^c_FF}), into the $\beta$-decay rate expression of Eq.~\eqref{eq:beta}, we obtain the generalized $\beta$-decay rate for arbitrary transitions:
\begin{multline}
    dw^c_{\beta^{\mp}} = \dfrac{G_F^2 |V_{ud}|^2}{2 \pi^3}(E_0 - E_e)^2 k E_e dE_e \dfrac{d\Omega_k}{4\pi} \dfrac{d\Omega_{\nu}}{4\pi} \dfrac{4\pi}{(2J_i+1)}\\
    \times \bigg \lbrace \Big( F^{\mp}(Z, E_e) \mp
    4\alpha Z a E_e \Big)
    \mathcal M^{0} + \mathcal{M}^{c}_{\mathrm{Tr}} + \mathcal{M}^{c}_{\mathrm{Op}} \bigg \rbrace,
    \label{eq:beta_final}
\end{multline}
where the explicit terms of $\mathcal M^{0}$, $ \mathcal{M}^{c}_{\mathrm{Tr}}$, and $\mathcal{M}^{c}_{\mathrm{Op}}$, are given in Eqs. \eqref{eq:M^0} , ~\eqref{eq:M^c_Tr}, and \eqref{eq:Op}, respectively.

\subsection {\textbf{The $\beta$-decay rates for allowed Gamow--Teller and unique first-forbidden transitions}}
At this final stage of the analysis, we consider the Gamow--Teller (GT) transition, characterized by 
$J_f^\pi - J_i^\pi = 1^+$, and the unique first-forbidden transition,
characterized by $J_f^\pi - J_i^\pi = 2^-$. Starting from the general $\beta$-decay formalism,
we evaluate the corresponding contributions and subsequently simplify the
expressions in the long-wavelength (LW) limit, $q r \ll 1$.
Nuclear-structure corrections beyond the LW approximation can be taken
directly from Refs.~\cite{GlickMagid2022, GlickMagid2023}.
\\

\textbf{Gamow-Teller transition}: We proceed by examining the allowed GT transition with $J=1$ in the LW limit, where the only nonvanishing nuclear multipoles satisfy
\begin{equation}
\hat{\mathcal{T}}^{el}_1(q)=\sqrt{2}\,\hat{\mathcal{L}}_1(q) + \hat{\mathcal T}^{\mathrm{el},(\mathrm{res})}_{1}.
\end{equation}
The operator $\hat{\mathcal L}_{1}$ provides the dominant leading-order contribution, while $\hat{\mathcal T}^{\mathrm{el},(\mathrm{res})}_{1}$ and the remaining multipoles enter only at subleading order in the LW expansion \cite{GlickMagid2022}. We then obtain
\begin{equation}
    \mathcal{M}^0_{(\mathrm{LW,\,GT})} = 3
    \bigg(1- \dfrac{1}{3}\hat{\boldsymbol{\nu}}\!\cdot\!\boldsymbol{\beta}\bigg)
    \big|\langle J_f \Vert \hat{\mathcal{L}}_{1}(q)\Vert J_i\rangle\big|^{2}.
\end{equation}
Furthermore, $\mathcal{M}^c_{\mathrm{Tr} (\mathrm{LW,\,GT})} = \mathcal{M}^c_{\mathrm{Op} (\mathrm{LW,\,GT})} = 0$ (we note that this is also the case for the LW limit of the allowed Fermi transition).

Substituting these results into the $\beta$-decay rate expression (Eq.~\eqref{eq:beta_final}), and using the definition of $\varepsilon_{EC}$ introduced in Eq.~\eqref{eq:psilon_EC}, with $a=R/\sqrt{6}$, we obtain
\begin{multline}
    dw^c_{\beta^{\mp} (\mathrm{LW,\,GT})}=
    \dfrac{G_F^2 |V_{ud}|^2}{2\pi^3}
    (E_0-E_e)^2\,k\,E_e\,dE_e\;
    \frac{d\Omega_k}{4\pi}\;
    \frac{d\Omega_\nu}{4\pi}\\
    \times \frac{4\pi}{2J_i+1} \Big|\langle J_f\Vert\hat{\mathcal{L}}_1(q)\Vert J_i\rangle\Big|^2
     3
    \Big(1- \dfrac{1}{3}\hat{\boldsymbol{\nu}} \cdot \boldsymbol{\beta}\Big) \\
    \times \bigg\{ F^{\mp}(Z,E_e)  \mp \dfrac{4}{\sqrt{6}}\varepsilon_{EC}
    \bigg\}
\end{multline}
\\

\textbf{Unique First-forbidden transition:} We now evaluate unique first-forbidden transitions, characterized by
$\Delta J^{\pi}=2^-$.
We simplify the $\beta$-decay expressions in the LW limit appropriate for this transition.
In this limit, the spherical Bessel functions scale as
$j_{2}(qr)\sim (qr)^2$ and $j_{1}(qr)\sim qr$.
Using the identity
\begin{equation}
\hat{\mathcal T}^{\mathrm{el}}_{2}
=
\sqrt{\frac{3}{2}}\,
\hat{\mathcal L}_{2}
+
\hat{\mathcal T}^{\mathrm{el},(\mathrm{res})}_{2},
\end{equation}
the electric multipole operator may, at leading order, be written
in terms of the longitudinal operator, which scales as
$\mathcal{O}(qr)$.
Accordingly, for unique $2^{-}$ transitions, the axial longitudinal
multipole $\hat{\mathcal L}_{2}$ provides the dominant leading-order
contribution, while
$\hat{\mathcal T}^{\mathrm{el},(\mathrm{res})}_{2}$ and the remaining
multipoles enter only at subleading order in the LW expansion.

Applying the LW limit to Eqs.~\eqref{eq:M^0}, \eqref{eq:M^c_Tr}, \eqref{eq:Op}, we obtain:
\begin{multline}
\mathcal{M}^0_{(\mathrm{LW}, \, 2^-)} =
\frac{5}{2}\bigg[
1
-\frac{2}{5}\,\hat{\boldsymbol{\nu}} \cdot \boldsymbol{\beta}
+\frac{1}{5}
(\hat{\boldsymbol{\nu}} \cdot \hat{\boldsymbol{q}})
(\boldsymbol{\beta} \cdot \hat{\boldsymbol{q}})
\bigg] \\
\times
\Big|\langle J_f \Vert \hat{\mathcal{L}}_{2}(q)\Vert J_i\rangle\Big|^{2},
\end{multline}
\begin{multline}
    \mathcal{M}^c_{\mathrm{Tr} (\mathrm{LW}, \, 2^-)}
    = \pm
    \alpha Z a\dfrac{ E_e^2}{q}\,\dfrac{16}{3} \\
    \times \bigg[
    (\hat{\boldsymbol{\nu}} \cdot \hat{\boldsymbol{q}})\beta^2
    +(\boldsymbol{\beta} \cdot \hat{\boldsymbol{q}})
    (\hat{\boldsymbol{\nu}} \cdot \boldsymbol{\beta})
    -2(\hat{\boldsymbol{\nu}} \cdot \hat{\boldsymbol{q}})
    (\boldsymbol{\beta} \cdot \hat{\boldsymbol{q}})^2
    \bigg] \\
    \times \frac{1}{2}\,
    \Big|\langle J_f\Vert \hat{\mathcal L}_{2}(q)\Vert J_i\rangle\Big|^2 ,
\end{multline}
and
\begin{multline}
    \mathcal{M}^c_{\mathrm{Op} (\mathrm{LW}, \, 2^-)}
    =
    \pm \alpha Z a \dfrac{ E_e^2}{q} \dfrac{16}{3}
    \big(\boldsymbol{\beta} \cdot \hat{\boldsymbol{q}}\big) 
    \\
    \times
    \left[
    1-\frac{2}{5}\,\hat{\boldsymbol{\nu}} \cdot \boldsymbol{\beta}
    +\frac{1}{5}(\hat{\boldsymbol{\nu}} \cdot \hat{\boldsymbol{q}})
    (\boldsymbol{\beta} \cdot \hat{\boldsymbol{q}})
    \right] \\
    \times \frac{5}{2}
    \Big|\langle J_f\Vert \hat{\mathcal L}_2(q)\Vert J_i\rangle\Big|^2.
\end{multline}
Substituting the above results into the general corrected $\beta$-decay expression, Eq.~\eqref{eq:beta_final}, and expressing the Coulomb correction in terms of $\varepsilon_{EC}$, we obtain the following LW-limit result for a unique first-forbidden transition:
\begin{align}
& dw^c_{\beta^{\mp} (\mathrm{LW}, \, 2^-)} \nonumber \\
& =
\dfrac{G_F^2 |V_{ud}|^2}{2 \pi^3}
\big(E_0 - E_e \big)^2 k E_e dE_e
\dfrac{d\Omega_k}{4\pi}
\dfrac{d\Omega_{\nu}}{4\pi}
\dfrac{4\pi}{2J_i+1}
\nonumber \\
&\times
\Big|\langle J_f \Vert \hat{\mathcal{L}}_{2}(q)\Vert J_i\rangle\Big|^{2} \bigg(F^{\mp}(Z, E_e)
\mp \dfrac{4}{\sqrt{6}} \varepsilon_{EC}\bigg)\nonumber \\
&\times
\frac{5}{2}\Bigg \lbrace
\bigg[
1 -\frac{2}{5}\,\hat{\boldsymbol{\nu}} \cdot \boldsymbol{\beta}
+\frac{1}{5}
\big(\hat{\boldsymbol{\nu}} \cdot \hat{\boldsymbol{q}}\big)
\big(\boldsymbol{\beta} \cdot \hat{\boldsymbol{q}}\big)
\bigg]
\nonumber \\
&\quad
\pm \dfrac{16}{3\sqrt{6}} \varepsilon_{EC}\dfrac{E_e}{q} 
\Big[\frac{1}{5}\big(\hat{\boldsymbol{\nu}} \cdot \hat{\boldsymbol{q}}\big)\beta^2
-  \frac{1}{5}\big(\boldsymbol{\beta} \cdot \hat{\boldsymbol{q}}\big)
\big(\hat{\boldsymbol{\nu}} \cdot \boldsymbol{\beta}\big)
\nonumber \\
&\qquad\qquad
+\big(\boldsymbol{\beta} \cdot \hat{\boldsymbol{q}}\big)  -  \frac{1}{5}\big(\hat{\boldsymbol{\nu}} \cdot \hat{\boldsymbol{q}}\big)
\big(\boldsymbol{\beta} \cdot \hat{\boldsymbol{q}}\big)^2
\Big]
\Bigg \rbrace 
\end{align}

\begin{table*}[t]
\caption{
Estimate of the maximal Coulomb correction 
$\varepsilon_{EC}^{\max}=\alpha Z\,E_e^{\max} R_{\mathrm{ch}}/(\hbar c)$, evaluated at 
$E_e^{\max}$ using $Q$ values from evaluated nuclear mass data~\cite{AME2020, Filianin2021Re187}. 
For $\beta^{-}$ decay we use $E_e^{\max}=Q+m_e$, while for $\beta^{+}$ decay we use 
$E_e^{\max}=Q-m_e$. 
Experimental rms charge radii $R_{\mathrm{ch}}$ are taken from 
Refs.~\cite{Angeli2013,Zhao2024ChargeRadii}. 
The last two columns, $\varepsilon_{EC}^{\max}\alpha Z$ and 
$\varepsilon_{EC}^{\max}E_e^{\max}R_{\mathrm{ch}}/(\hbar c)$, reflect the order of magnitude of 
higher-order corrections relative to $\varepsilon_{EC}^{\max}$.
}
\label{tab:epsilon_EC_max_reorg}
\centering

\renewcommand{\arraystretch}{1.4}
\setlength{\tabcolsep}{9pt}

\begin{tabular}{c c c c c c}
\hline
Nucleus & $E_e^{\max}$ [MeV] & $R_{\mathrm{ch}}$ [fm] & $\varepsilon_{EC}^{\max}$ & $\varepsilon_{EC}^{\max}\alpha Z$ & $\varepsilon_{EC}^{\max}E_e^{\max}R_{\mathrm{ch}}/(\hbar c)$ \\
\hline

$^{3}_{1}\mathrm{H}$ & 0.52959 & 1.7591 & $3.45\times10^{-5}$ & $2.51\times10^{-7}$ & $1.63\times10^{-7}$ \\

$^{6}_{2}\mathrm{He}$ & 4.02 & 2.0660 & $6.14\times10^{-4}$ & $8.97\times10^{-6}$ & $2.59\times10^{-5}$ \\

$^{8}_{3}\mathrm{Li}$ & 16.52 & 2.3390 & $4.29\times10^{-3}$ & $9.38\times10^{-5}$ & $8.39\times10^{-4}$ \\

$^{11}_{6}\mathrm{C}$ & 1.47 & 2.32 & $7.57\times10^{-4}$ & $3.31\times10^{-5}$ & $1.31\times10^{-5}$ \\

$^{15}_{8}\mathrm{O}$ & 2.24 & 2.69 & $1.78\times10^{-3}$ & $1.04\times10^{-4}$ & $5.44\times10^{-5}$ \\

$^{23}_{10}\mathrm{Ne}$ & 4.89 & 2.9104 & $5.26\times10^{-3}$ & $3.84\times10^{-4}$ & $3.80\times10^{-4}$ \\

$^{32}_{18}\mathrm{Ar}$ & 10.62 & 3.3468 & $2.37\times10^{-2}$ & $3.11\times10^{-3}$ & $4.26\times10^{-3}$ \\

$^{38}_{19}\mathrm{K}$ & 5.40 & 3.4264 & $1.30\times10^{-2}$ & $1.80\times10^{-3}$ & $1.22\times10^{-3}$ \\

$^{187}_{75}\mathrm{Re}$ & 0.513 & 5.3596 & $7.64\times10^{-3}$ & $4.18\times10^{-3}$ & $1.07\times10^{-4}$ \\

\hline
\end{tabular}
\end{table*}

\section{\textbf{Discussion and Conclusions}}
We developed a systematic model-independent framework for nuclear $\beta$ decay that incorporates the leading nuclear-structure electrostatic correction in arbitrary angular momentum transitions. Coulomb effects were treated through single-photon exchange between the emitted charged lepton and the nuclear charge distribution, resulting in a Coulomb-distorted lepton wave function and a consistent perturbative expansion of the decay amplitude in both the exchanged-photon momentum and nuclear multipole operators.

Using this framework, we derived the leading finite-size Coulomb correction to the decay rate and demonstrated the formalism explicitly for Gamow--Teller and unique first-forbidden transitions in the long-wavelength limit. The correction originates from the momentum shift induced by the exchanged Coulomb photon and enters the electrostatically corrected decay-rate expression through the nuclear charge form factor, the leptonic current, and the Coulomb-induced shift of the momentum transfer in the multipole operators.

The formalism naturally leads to a decoupling of the Fermi function $F(Z,E_e)$, which accounts for the point-nucleus Coulomb interaction to all orders in $\alpha Z/k$, and the nuclear-structure-dependent electrostatic contribution appears at order
$\varepsilon_{EC}= \alpha Z\,E_e R$. Importantly, this contribution is obtained independently of a specific nuclear-structure model and therefore provides a general correction that can be incorporated into different nuclear many-body calculations.

Using the maximal electron energy $E_e^{\max}$, the representative estimates summarized in Table~\ref{tab:epsilon_EC_max_reorg} show that $\varepsilon_{EC}$ ranges from $\sim10^{-5}$ in very light nuclei to $\sim10^{-2}$ in medium-$Z$ nuclei relevant for precision $\beta$-decay studies. Higher-order terms, scaling as $(\alpha Z)\varepsilon_{EC}$ or $(E_eR)\varepsilon_{EC}$, are additionally suppressed and are expected to contribute at the relative level $\sim10^{-1}$--$10^{-3}$ with respect to the leading contributions presented in the Table. 
Consequently, these estimated uncertainties suggest that for many of the experimentally relevant nuclei, additional nuclear-structure Coulomb corrections remain small.

\section{\textbf{Acknowledgments}}
The authors wish to express their sincere thanks to Leendert Hayen for pointing them to key articles and studies on beta decay, which greatly supported and informed this research. This work was partially supported by the Israel Science Foundation grant no. 889/23.

\bibliographystyle{apsrev4-2}
\bibliography{references}

\end{document}